\documentclass[twocolumn,floatfix,aps,superscriptaddress,nofootinbib]{revtex4-1}
\usepackage{graphicx}
\usepackage{amsmath}
\usepackage{amssymb}
\usepackage{bm}

\begin{document}
	\title{Synthetic Altermagnets}
	
	\author{Ali Asgharpour}
	%\email{a.asgharpour@uu.nl}
	\affiliation{Institute for Theoretical Physics and Center for Extreme Matter and Emergent Phenomena, Utrecht University, Leuvenlaan 4, 3584 CE Utrecht, The Netherlands}
	
	\author{Bert Koopmans}
	%\email{b.koopmans@tue.nl}
	\affiliation{Department of Applied Physics, Center for NanoMaterials (cNM) Eindhoven University of Technology, P.O. Box 513, 5600 MB Eindhoven, The Netherlands}
	
	\author{Rembert A. Duine}
	\email{r.a.duine@uu.nl}
	\affiliation{Institute for Theoretical Physics and Center for Extreme Matter and Emergent Phenomena, Utrecht University, Leuvenlaan 4, 3584 CE Utrecht, The Netherlands}
	\affiliation{Department of Applied Physics, Center for NanoMaterials (cNM) Eindhoven University of Technology, P.O. Box 513, 5600 MB Eindhoven, The Netherlands}
	
	\date{\today}% It is always \today, today,
	%  but any date may be explicitly specified
	
	\begin{abstract}
		
		Altermagnets, a distinct class of antiferromagnets with electronic structures resembling those of $d$-wave superconductors, exhibit intriguing properties that have gained significant attention in recent research. In this article, we propose synthetic altermagnets, composed of two anisotropic ferromagnetic layers arranged such that the total net magnetization is zero. We investigate the properties of these synthetic altermagnets, focusing on their electronic band structures, spin current, Berry curvature, and anomalous Hall conductivity. By developing a minimal model for our synthetic altermagnets, we examine the influence of factors such as anisotropic coupling strengths and spin-orbit coupling on the physical phenomena altermagnets manifest. Our findings open a new path for the realization of altermagnetic materials experimentally and highlight their potential applications in magneto-electronics, magneto-optics, and spintronics devices.
		
	\end{abstract}
	
	\pacs{Valid PACS appear here}
	\maketitle
	
	\section{Introduction}
	\label{sec_intro}
	
	There has recently been considerable focus on altermagnets, a distinctive class of antiferromagnets characterized by their electronic structure, which resembles that of $d$-wave superconductors or superconductors with higher angular momentum Cooper pairs~\cite{Sinova,phase,magnon}. This structural similarity not only underscores the intrinsic importance of altermagnets but also highlights their potential for diverse applications~\cite{spin-splitter,GMR,tiltedspincurrent,observ.SST,observ.SST2,spin-neutral-current,topoSC,Josephson,AndrevReflection, AndrevReflection2,Majorana,thermaltransport,Cooper}. Antiferromagnets inherently possess advantages over ferromagnets, such as robustness against external perturbations and rapid dynamic responses. The unique electronic band structure of altermagnets further presents intriguing possibilities for accessing and manipulating spin and charge transport properties.
	
	In ferromagnets, the spin order is uncompensated in the momentum space of electrons, with distinct nondegenerate spin-up and spin-down channels resulting in a nonzero net magnetization. In contrast, antiferromagnets exhibit no such spin order in momentum space, complicating their detection and manipulation. Altermagnetism is characterized by a compensated spin order in electronic momentum space, featuring nondegenerate and anisotropic spin-up and spin-down channels that are symmetrically interrelated through rotations in real space. This configuration suggests that altermagnetism could be the magnetic analogue of $d$-wave superconductors. In spintronics, altermagnets offer scalability comparable to antiferromagnets in both spatial and temporal dimensions, while also providing efficient methods for detecting and controlling magnetic states, similar to ferromagnets.
	
	Altermagnets demonstrate a compensated magnetic structure in real space, with sublattices of opposite spins connected through crystal-rotation symmetries. Concurrently, they exhibit an unconventional spin-polarization arrangement in reciprocal momentum space that reflects these rotational symmetries. This interplay between real and reciprocal space results in electronic band configurations with broken time-reversal symmetry and momentum-dependent alternating spin splitting~\cite{DFTbadnsplittingMnO2,spinsplitting,DFTd-waveRuO2,predictionspin-splitting,spinsplittingMnF2,spinsplittingDFTMnF2,TRSbreaking-observationAHE,spin-splitting,spinsplittingmagneticgroups}. In addition to the first-principles calculations of unconventional bandstructure and spin-splitting of altermagnets leading to spin current, experimental observations have also confirmed these properties~\cite{liftedKramersEx,Observ.spinsplitting,ObservSS,ObservSS2,ObservSS3,observ.bands}. Furthermore, like ferromagnets, anomalous Hall effect (AHE) is another distinctive feature of altermagnets, arising from the broken time-reversal and inversion symmetries in these materials. This property differentiates altermagnets from conventional antiferromagnets. The emergence of AHE has been both theoretically predicted and experimentally validated~~\cite{SHEDFTRuO2,TRSbreaking-observationAHE,thermaltransport,AHEinRuO2,AHCCoNbS,AHCorganic,AHCperovskites,AHCsemiconductor,AHCCaCrO,AHC2mat,AHCDirac,AHCrutile,AHCMnSi}.

	\begin{figure}[h!]
		\centerline{\includegraphics[width=0.98\linewidth]{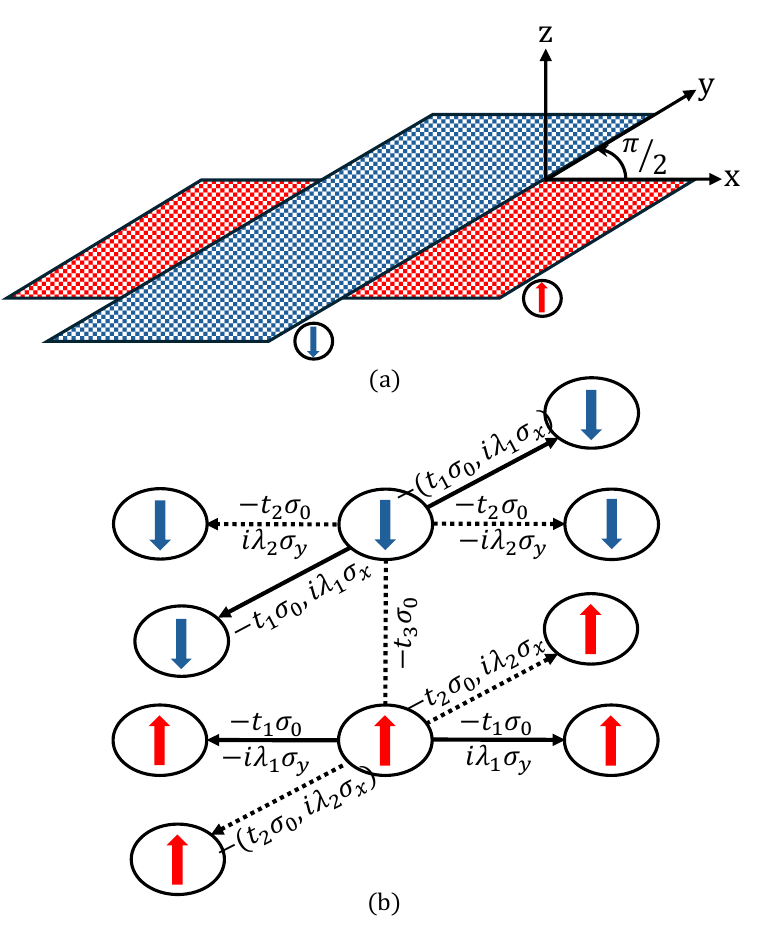}}
		\caption{(a) An illustration depicting two ferromagnetic layers with opposite magnetization, where one layer is rotated by $\pi/2$ relative to the other. (b) The diagram presents the tight-binding hopping parameters both within each layer and between the layers. Solid and dashed lines indicate different hopping parameters between lattice sites.}
		\label{fig:system}
	\end{figure}

	Altermagnetic materials are typically found within the rutile crystal family~\cite{spin-splitter,GMR,spin-neutral-current,SHEDFTRuO2,AHEinRuO2,Observ.spinsplitting,observ.bands}, but there has been much ongoing research on other candidate materials~\cite{DFTbadnsplittingMnO2,MnF2theory,Perovskitetheory,LaMO3DFT,dopedFeSb2theory,Spin-split collinear antiferromagnets: A large-scale DFT,spinsplittingDFTMnF2,liftedKramersEx,ObservSS,ObservSS2,ObservSS3}. Historical studies, including those by Néel, recognized the compensated antiparallel alignment of magnetic moments in several insulating rutile compounds as a classic manifestation of antiferromagnetism. However, these early studies primarily focused on the magnetic lattice and overlooked the critical influence of nonmagnetic atoms within rutile crystals on their magnetic properties. This oversight may explain why the unconventional spin polarization order in momentum space, leading to time-reversal symmetry disruption and alternating spin splitting in their nonrelativistic band structures, remained undetected for nearly a century. Systematic classification based on spin-symmetry principles now distinguishes the altermagnetic phase from other magnetic phases~\cite{Sinova}. Key criteria for classifying a material as an altermagnet include having an even number of magnetic atoms in the unit cell, a total net magnetization of zero, no inversion center between magnetic atoms with opposite spins, and rotational or combined translational and inversion symmetry connecting the opposite-spin sublattices. Recent studies aimed at developing minimal models to capture various characteristics of altermagnets have paved the way for further research in this area~\cite{Arne, models}.
	
	In this work, we propose a novel approach to experimentally realize altermagnets that in principle has the advantage of relying on ferromagnetic materials of which e.g. the domain structure is well understood. Our approach involves constructing synthetic altermagnets from two ferromagnetic layers rotated by $\pi/2$. Each layer is designed to exhibit an anisotropic electronic band structure. This anisotropy could be either due to intrinsic anisotropy, or could result from the growth direction of the layers. We develop a straightforward tight-binding model for these synthetic altermagnets and investigate their electronic properties, as well as their AHE, which serves as a distinguishing measure between altermagnets and antiferromagnets. Figure~\ref{fig:system} illustrates our proposed synthetic altermagnet system. To break inversion symmetry in each layer, we need either an anisotropic ordering of local orbitals~\cite{anisotropicOrbitalOrdering} or the effect of nonmagnetic atoms~\cite{nonmagneticeffectonspinsplitting}. In our description, both of these are modelled by anisotropic hopping parameters. 
	
	Our paper is organized as follows: In Section~\ref{sec_model}, we introduce the minimal model of our system, and explore the electronic properties of our model. In section~\ref{sec_SC}, we calculate spin current and polarization for our model using Boltzmann transport equation. Next, in Section~\ref{sec_AHE}, we incorporate spin-orbit coupling and compute the Berry curvature and anomalous Hall conductivity (AHC) for our system in the presence of inter-layer and intra-layer Rashba couplings. Finally, we present our concluding remarks in Section~\ref{sec_conclusions}.

	\section{Model}
	\label{sec_model}
	
	We propose a system comprising two ferromagnetic layers, arranged in such a way that their net magnetization is zero. The layers are designed to be transformable into one another through a $\pi/2$ rotation, thereby satisfying the conditions necessary for altermagnetism. The unit cell of this system includes magnetic atoms with anisotropic ordering of local orbitals. Alternatively, nonmagnetic atoms can also be considered, but in our model, we primarily focus on the variations in hopping parameters as a result of these mechanisms, which ultimately induce an altermagnetic state. To further refine our model, we incorporate spin-orbit coupling (SOC) both within each layer and between layers. Specifically, we focus on the Rashba SOC in our minimal model. The intra-layer Rashba Hamiltonian is given by: ${H^{\text{intra}}_R =  i \lambda_R \sum_{\langle i,j\rangle} c_i^{\dagger} (\boldsymbol{\sigma} \times \hat{\mathbf{d}}_{ij})\cdot \hat{\mathbf{z}} c_j }$ where $\hat{\mathbf{d}}_{ij}$ is the unit vector along the bond through which the electron moves from site $j$ to site $i$. The Rashba Hamiltonian in each layer explicitly violates the $z \rightarrow -z$ mirror symmetry~\cite{Kane}. As we discuss in Sec.~\ref{sec_AHE} in more detail, to get a nonzero anomalous Hall effect, we need to introduce an inter-layer Rashba SOC that results from a tilted electric field.
	
	We neglect in first instance SOC, and we will include it later, therefore, the tight-binding model that describes electronic properties of our system is as follows,
	\begin{eqnarray}
		H_e =& -&t_1 \left(\sum_{\langle i,j\rangle_x, \sigma} c^{\dagger}_{1,i, \sigma} c_{1,j,\sigma} + \sum_{\langle i,j\rangle_y, \sigma} c^{\dagger}_{2,i, \sigma} c_{2,j,\sigma} \right) \nonumber \\
		&-& t_2 \left(\sum_{\langle i,j\rangle_y, \sigma} c^{\dagger}_{1,i, \sigma} c_{1,j,\sigma} + \sum_{\langle i,j\rangle_x, \sigma} c^{\dagger}_{2,i, \sigma} c_{2,j,\sigma} \right) \nonumber\\
		&-& \mu \sum_{m,i, \sigma} c^{\dagger}_{m,i, \sigma} c_{m,i,\sigma} - t_3 \sum_{i, \sigma} c^{\dagger}_{1,i, \sigma} c_{2,i,\sigma} \nonumber \\&-& J_{sd} \sum_{m,i, \sigma, \sigma'} \mathbf{S}_{m,i} \cdot c^{\dagger}_{m,i, \sigma} \bm{\sigma}_{\sigma \sigma'} c_{m,i,\sigma'},
	\end{eqnarray}

	\noindent where $c^{(\dagger)}_{m,i, \sigma}$ is the (creation) annihilation operator of an electron at site $i$ in layer $m$ with spin $\sigma$. The notation $\langle i,j\rangle_{x(y)}$ signifies nearest-neighbour hopping in $x(y)$- direction. The parameters $t_1$ and $t_2$ indicate different intra-layer coupling strengths along each layer's directions, influenced by either the inclusion of vacuum sites within the unit cell or the anisotropic ordering of local orbitals. The parameter $t_3$ denotes the inter-layer coupling strength, where layers have opposite magnetization. The chemical potential $\mu$ determines the overall doping level of the system. The coupling $J_{sd}$ quantifies the onsite exchange interaction between the spin of the itinerant electrons and the localized spins on the sites $\mathbf{S}_{m,i}$. For simplicity, we use $\mathbf{S}_{m,i} = (-1)^{m-1}S \hat{z}$, with $m=1$ and $m=2$ corresponds to the bottom layer and top layer, respectively, as shown in Fig.~\ref{fig:system}.
	
	Neglecting spin-orbit coupling (SOC), our Hamiltonian model remains block diagonal in spin space. To facilitate analysis, we perform a Fourier transform and rewrite the operators in terms of electron band operators ${d_{n,\mathbf{k},\sigma} = \sum_{m} q^*_{n,m,\mathbf{k},\sigma} c_{m,\mathbf{k},\sigma}}$ with $q^*_{n,m,\mathbf{k}, \sigma}$ is chosen such that
	\begin{equation}
		H_e = \sum_{n,\mathbf{k},\sigma} E_{n,\mathbf{k},\sigma} d^{\dagger}_{n,\mathbf{k},\sigma} d_{n,\mathbf{k},\sigma}.
	\end{equation}
	\begin{figure}[t!]
		\centerline{\includegraphics[width=0.98\linewidth]{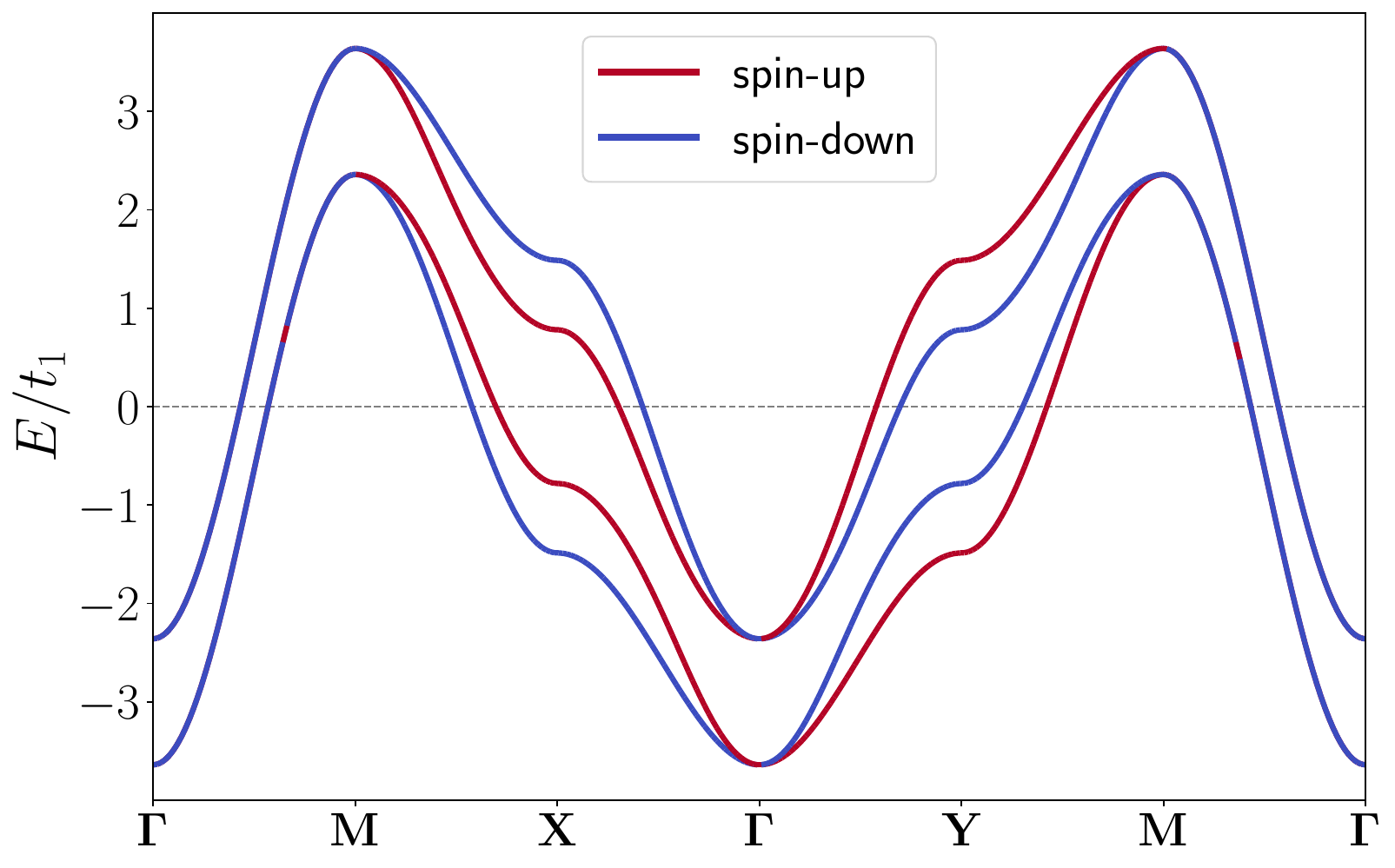}}
		\caption{The electron dispersion in the $x$- and $y$-direction. The red and blue colors denote spin-up and spin-down, respectively. We used the parameters, $t_2/t_1 =0.5$, $t_3/t_1 =0.5$, $\mu/t_1 =0$, $J_{sd}S/t_1 =0.4$.}
		\label{fig:disp1}
	\end{figure}
	Figure~\ref{fig:disp1} illustrates the electronic spectrum of our synthetic altermagnet. The bands display the characteristic spin splitting associated with altermagnets, exhibiting a $d$-wave symmetry. If the hopping coefficients in the $x$ and $y$ directions are equal, the spin-split bands disappear, and the system behaves as a conventional antiferromagnet, as anticipated.
	
\begin{figure}[t!]
		\centerline{\includegraphics[width=0.98\columnwidth]{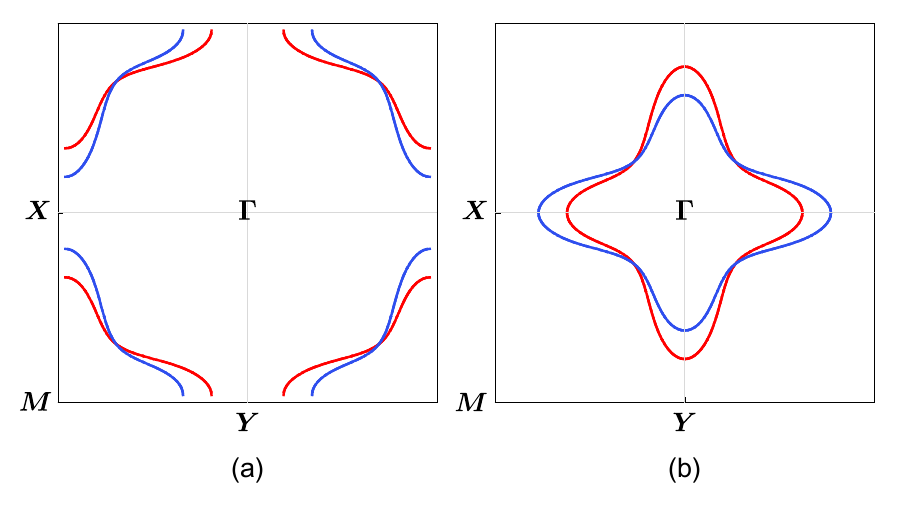}}
		\caption{The Fermi surface at two distinct doping levels. Red and blue correspond to spin-up and spin-down polarization, respectively. The plots are for $t_2/t_1 =0.5$, $t_3/t_1 =0.5$, $J_{sd}S/t_1 =0.4$, and (a) $\mu/t_1 =2$, (b) $\mu/t_1 =-2$.}
		\label{fig:FS}
	\end{figure}

	Figure~\ref{fig:FS} depicts the Fermi surfaces at two doping levels, corresponding to the electronic spectrum shown in Fig.~\ref{fig:disp1}. The spin-degenerate points along the diagonal 
	$k_x = \pm k_y$ originate from the mirror symmetry inherent in our system. These points are analogous to the Cooper-pair nodes found in unconventional $d$-wave superconductors. The intra-layer coupling term, $t_3$, induces a gap between bands with the same spin in each layer.
	
	The minimum magnitude of the unconventional spin-splitting depends on the tight-binding parameters, and can be expressed in momentum space as follows,
	\begin{flalign}
		|\Delta^{ss}_{e}| &=  \sqrt{(J_{sd} S + (t_1-t_2)(\cos(k_x a) - \cos(k_y a)))^2 + t_3^2} \nonumber\\
		& - \sqrt{(J_{sd} S - (t_1-t_2)(\cos(k_x a) - \cos(k_y a)))^2 + t_3^2}.
		\label{eq:gap}
	\end{flalign}
	This equation highlights how the hopping parameters and the onsite exchange interaction strength influence the spin-splitting in the electronic band structure. When there are no anisotropic hopping terms in different directions, the band structure does not exhibit unconventional spin splitting, resulting in a typical antiferromagnetic band structure as expected. Furthermore, the anisotropic ordering of orbitals induces different spin-spin exchange coupling and the unconventional splitting magnon bands~\cite{chiralmagnonsRuO2}, see Appendix~\ref{app_magnon}.

\begin{figure}[t!]
		\centerline{\includegraphics[width=0.98\linewidth]{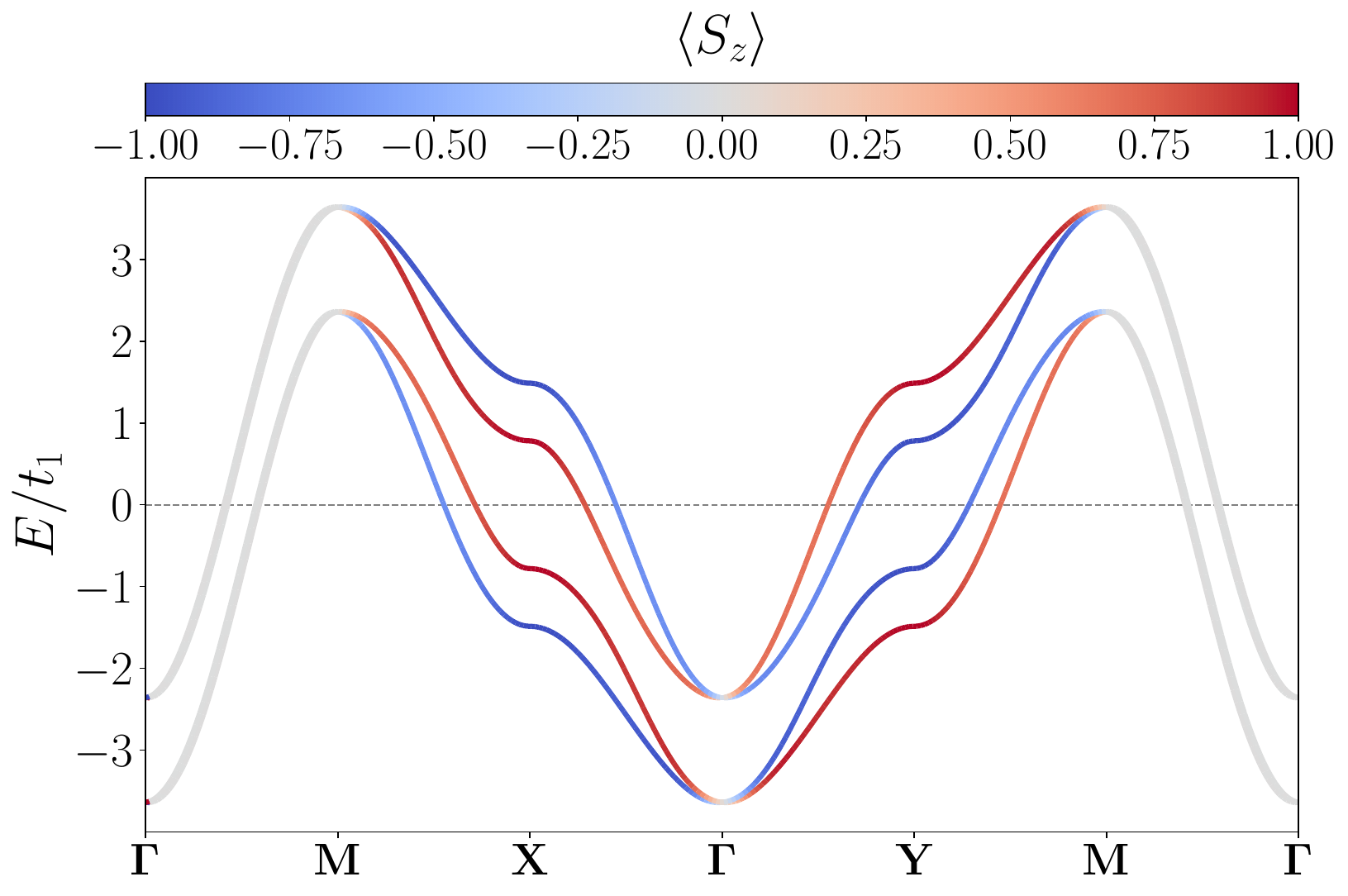}}
		\caption{The band dispersion for our system with intra-layer Rashba coupling, but without inter-layer Rashba coupling, and the following parameters: $t_2/t_1 =0.5$, $t_3/t_1 =0.5$, $J_{sd}S/t_1 =0.4$, $\lambda_1/t_1 =0.2$, $\lambda_2/t_1=0.1$, and $\mu/t_1=0.0$. The color denotes the expectation value of spin in $z$-direction.}
		\label{fig:dispSOC}
	\end{figure}

	Now, we add the intra-layer Rashba SOC Hamiltonian to our system. We use $\pm\hat{\mathbf{z}}$ for different layers according to their normal plane vectors. Therefore, the tight-binding model of the intra-layer Rashba Hamiltonian is described as follows in our system:

	\begin{flalign}
		H^{\text{intra}}_R &=  i  \left(\sum_{\langle i,j\rangle_x, \sigma} -\lambda_1\sigma_y c^{\dagger}_{1,i, \sigma} c_{1,j,\sigma}+ \lambda_2\sigma_y c^{\dagger}_{2,i, \sigma} c_{2,j,\sigma}\right) \nonumber \\
		&+ i  \left(\sum_{\langle i,j\rangle_y, \sigma} \lambda_2\sigma_x c^{\dagger}_{1,i, \sigma} c_{1,j,\sigma}  -\lambda_1\sigma_x c^{\dagger}_{2,i, \sigma} c_{2,j,\sigma} \right),
	\end{flalign}
	where $\lambda_{1(2)}$ denotes different intra-layer Rashba coupling strength originating from distinct hopping parameter in various directions. Figure~\ref{fig:dispSOC} represents the electron bands dispersion of our model with considering the Rashba SOC with anisotropic strength in different directions within each layer. The introduction of SOC results in spin-mixed states, and alters the expectation value of the spin in the $z$-direction of the band structure as shown in Fig.~\ref{fig:dispSOC} as compared to Fig.~\ref{fig:disp1}, where SOC is absent.

	\section{Spin Current and Polarization}
	\label{sec_SC}
	
	The spin-polarized current is a distinctive characteristic of altermagnets, setting them apart from antiferromagnets due to their unconventional $d$-wave symmetric band structure. This feature opens up possibilities for spintronic applications in materials that exhibit no net magnetization. In this section, we calculate the spin current and polarization using the Boltzmann transport equation within the framework of our model. For simplicity, we exclude SOC terms to avoid complications arising from spin mixing. The $z$-component of spin current density at position $\mathbf{r}$ is expressed as
	\begin{equation}
		\mathbf{j}^{z}(\mathbf{r}, t) = \frac{\hbar}{2} \sum_{n,s} s \int_{\mathbf{\Omega}} \frac{d^2k}{(2 \pi)^2} f_{n,s} (\mathbf{r}, \mathbf{k}, t) \mathbf{v}_{n,s}(\mathbf{k}),
	\end{equation}
	where $s=\pm$ corresponds to spin-up and spin-down states, respectively. The integral is performed over the first Brillouin zone, denoted by $\Omega$. Here, $f_{n,s}$ represents the distribution function of the band with index $n$ and spin $s$, and $\mathbf{v}_{n,s} = 1/\hbar \nabla_{\mathbf{k}} E_{n,s} (\mathbf{k})$ denotes the group velocity associated with the band dispersion relation $E_{n,s}$.
	
	The distribution function is a solution of Boltzmann transport equation, and can be expressed as $f^0_{n,s} (\mathbf{r}, \mathbf{k}, t) + \delta f_{n,s}(\mathbf{r}, \mathbf{k},t)$, where $f^0_{n,s}$ is the equilibrium Fermi-Dirac distribution, and $\delta f_{n,s}$ represents the deviation from local equilibrium. We assume a constant temperature $T$, a spatially uniform $\boldsymbol{\mathcal{E}} = - \nabla(\phi - \mu/e)$, which is the gradient of the electrochemical potential, and a constant momentum relaxation time $\tau$ for both spin states. Under these conditions, the deviation $\delta f_{n,s}$ for steady states, derived from the linearized Boltzmann equation, is given by
	\begin{equation}
		\delta f_{n,s} = e \tau \boldsymbol{\mathcal{E}} \cdot \mathbf{v}_{n,s} \frac{\partial f^0_{n,s}}{\partial E_{n,s}}.
	\end{equation} 
	The spin conductivity tensor is defined by the linear relation between the spin current density and the electric field components, $j^z_{\alpha} = \sigma^z_{\alpha \beta} \mathcal{E}_{\beta}$. Consequently, we obtain
	\begin{equation}
		\sigma^z_{\alpha, \beta} = \frac{\hbar e \tau}{2} \sum_{n,s} s \int_{\mathbf{\Omega}} \frac{d^2k}{(2 \pi)^2} v_{n,s}^{\alpha} (\mathbf{k}) v_{n,s}^{\beta} (\mathbf{k}) \frac{\partial f^0_{n,s}}{\partial E_{n,s}}.
	\end{equation} 
	The integral yields a nonzero result only when $\alpha = \beta$, as the integrand becomes an odd function of 
	$\mathbf{k}$ for off-diagonal components, leading to cancellation.
	
	To quantify the degree of spin polarization in the system, we employ a polarization formula that compares the relative contributions of spin-up and spin-down electrons to the total current. The spin polarization along the $x$- or $y$- direction, current direction, is given by 
	\begin{equation}
		P_{x(y)} = \left(\frac{j_{\uparrow} - j_{\downarrow} }{j_{\uparrow} + j_{\downarrow} } \right)_{x(y)},
	\end{equation}
	where $j_{\uparrow}$ and $j_{\downarrow}$ denote the current contributions from spin-up, $s=+$, and spin-down, $s=-$, electrons, respectively. 
	\begin{figure}[t!]
		\centerline{\includegraphics[width=0.98\linewidth]{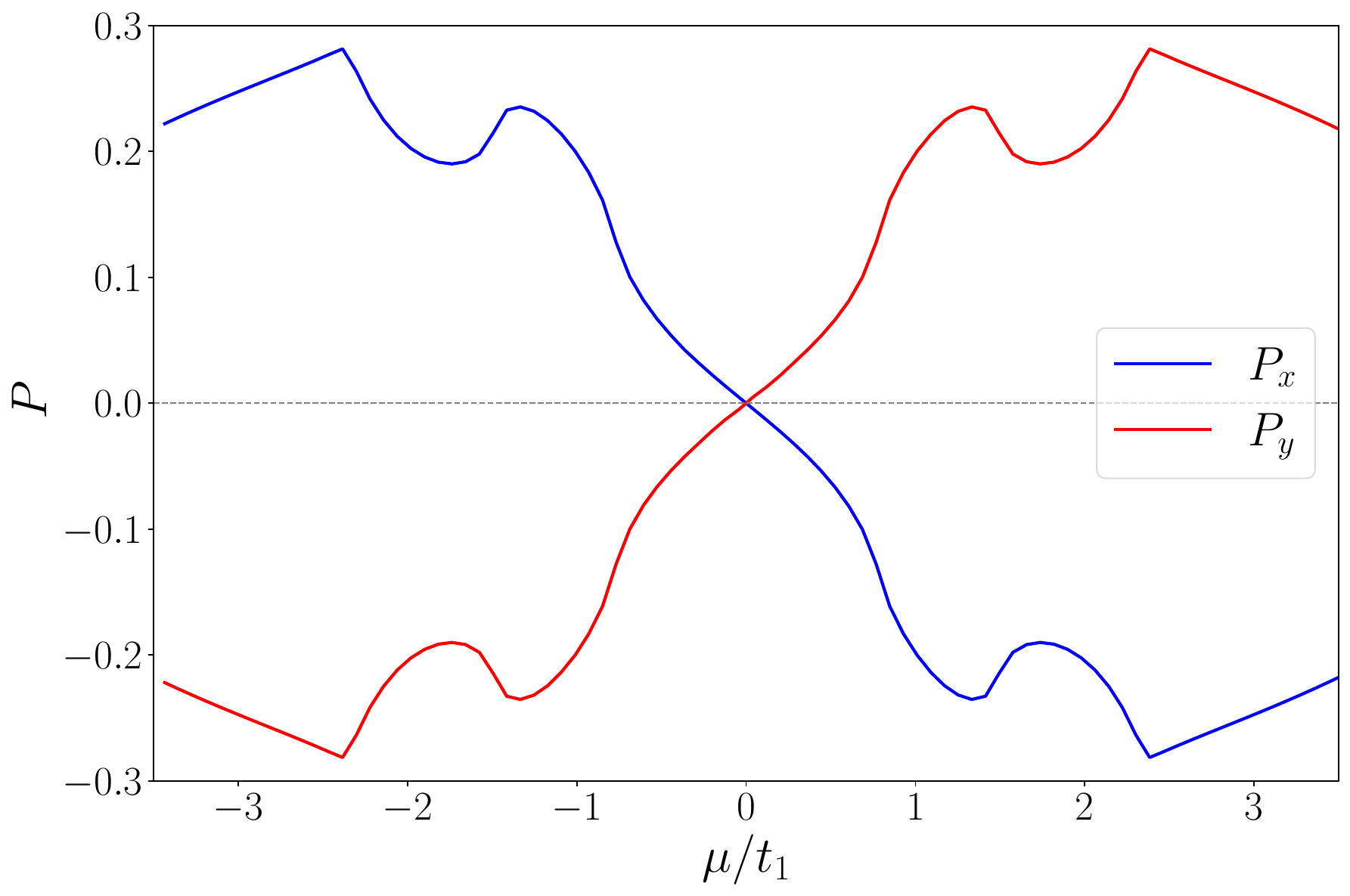}}
		\caption{The spin polarization along the $x$ and $y$ directions as a function of the dimensionless chemical potential ($\mu/t$) for our system without considering SOC terms. We use the following parameters: $t_2/t_1 =0.5$, $t_3/t_1 =0.5$, ${J_{sd}S/t_1 =0.4}$, and $k_B T / t_1 = 0.01$.}
		\label{fig:polarization}
	\end{figure} 
	Figure~\ref{fig:polarization} illustrates spin polarization along $x$ or $y$ current direction as a function of the chemical potential $\mu/t$ with $t$ the leading hopping parameter. The plot shows how the spin polarization varies as the chemical potential changes, indicating different regimes of spin-polarized behaviour. This variation is related to the underlying electronic structure of our model. We can control the spin-polarized current by changing the direction of the electric field or adjusting the chemical potential. Hence, our synthetic altermagnet proposal could be used as a spin valve in spintronic devices.
	
	\section{Berry Curvature And Anomalous Hall Effect}
	\label{sec_AHE}
	
	Altermagnets are predicted to exhibit an AHE due to spin-splitting and broken time-reversal symmetry in their electronic band structure. This prediction arises from theoretical models indicating that the alternating magnetic moments in altermagnets generate a non-zero Berry curvature in their electronic band structure, which is a crucial ingredient for AHE, and validated experimentally. Consequently, altermagnets possess a non-zero AHC, distinguishing them from antiferromagnets and aligning them more closely with ferromagnets in this regard. Investigating AHE in altermagnets is essential because it not only confirms these theoretical predictions but also opens up potential applications in spintronics, where controlling spin currents without external magnetic fields can lead to more efficient and advanced magnetic memory and logic devices. In this section, we investigate the Berry curvature and AHC for our system in the presence of SOC. The Rashba SOC leads to a modification of the electronic band structure as we presented in the previous section. This interaction can significantly impact the Berry curvature, thereby influencing the AHC.
	
	To compute the Berry curvature, we utilize the following expression~\cite{BerryPhase},
	\begin{equation}
		\boldsymbol{\Omega}_n =i \left(\sum_{n'\neq n}\frac{\left\{\langle n|\frac{\partial H(\mathbf{k})}{\partial{k_x}} | n' \rangle \langle n'|\frac{\partial H(\mathbf{k})}{\partial{k_y}} | n \rangle \right\}- \left\{ x \leftrightarrow y\right\}}{(E_n (\mathbf{k})- E_{n'}(\mathbf{k}))^2}\right)
	\end{equation}
	with $| n \rangle$ the periodic part of the Bloch wave function, and $E_{n} (\mathbf{k})$ the energy of the Bloch state for band $n$ with crystal momentum $k$. Furthermore, we numerically evaluate the AHC using the formula,
	\begin{equation}
		\sigma_{xy} =- \frac{e^2}{h} \int_{-\pi}^{\pi} \int_{-\pi}^{\pi} \frac{dk_x dk_y}{2 \pi} \sum_n f_{\text{FD}}(E_n) \boldsymbol{\Omega}_n,
	\end{equation}
	where $f_{\text{FD}}(E_n)$ is the Fermi-Dirac distribution function.  When we neglect the inter-layer coupling terms ($t_3 = 0$ and $\lambda_3 =0$), the Hamiltonian can be written in the block diagonal form, with each block corresponding to different layers. Consequently, we have two decomposed ferromagnetic layers with opposite magnetization, leading to the following expressions for the Berry curvatures:
	\begin{equation}
		\Omega_{1(2), \mp} = \mp(\pm) \frac{2 J_{sd}S \lambda_1 \lambda_2 \cos(k_x a) \cos(k_y a)}{F_{1(2)}^{3/2}},
	\end{equation}
	where the first and second indices of $\Omega$ denote different layers and corresponding bands, respectively, and ${F_{1(2)} = (J_{sd}S)^2 + 4(\lambda_{1(2)}^2 \sin^2(k_x a) + \lambda_{2(1)}^2 \sin^2(k_y a))}$. Although we find nonzero AHC for each layer as we expected for any ferromagnets with considering SOC, the total AHC of our system is zero for any chemical potential since AHC of each layer cancels the other one stemming from the different signs of Berry curvature for different layers. When we consider non-zero inter-layer coupling coefficient, $t_3 \neq 0$, and neglect inter-layer SOC, $\lambda_3 =0$, the Berry curvature corresponding to each band is more complicated. The total AHC remains zero because the inter-layer coupling term does not mix the spins of different layers. 
	\begin{figure}[t!]
		\centerline{\includegraphics[width=0.98\linewidth]{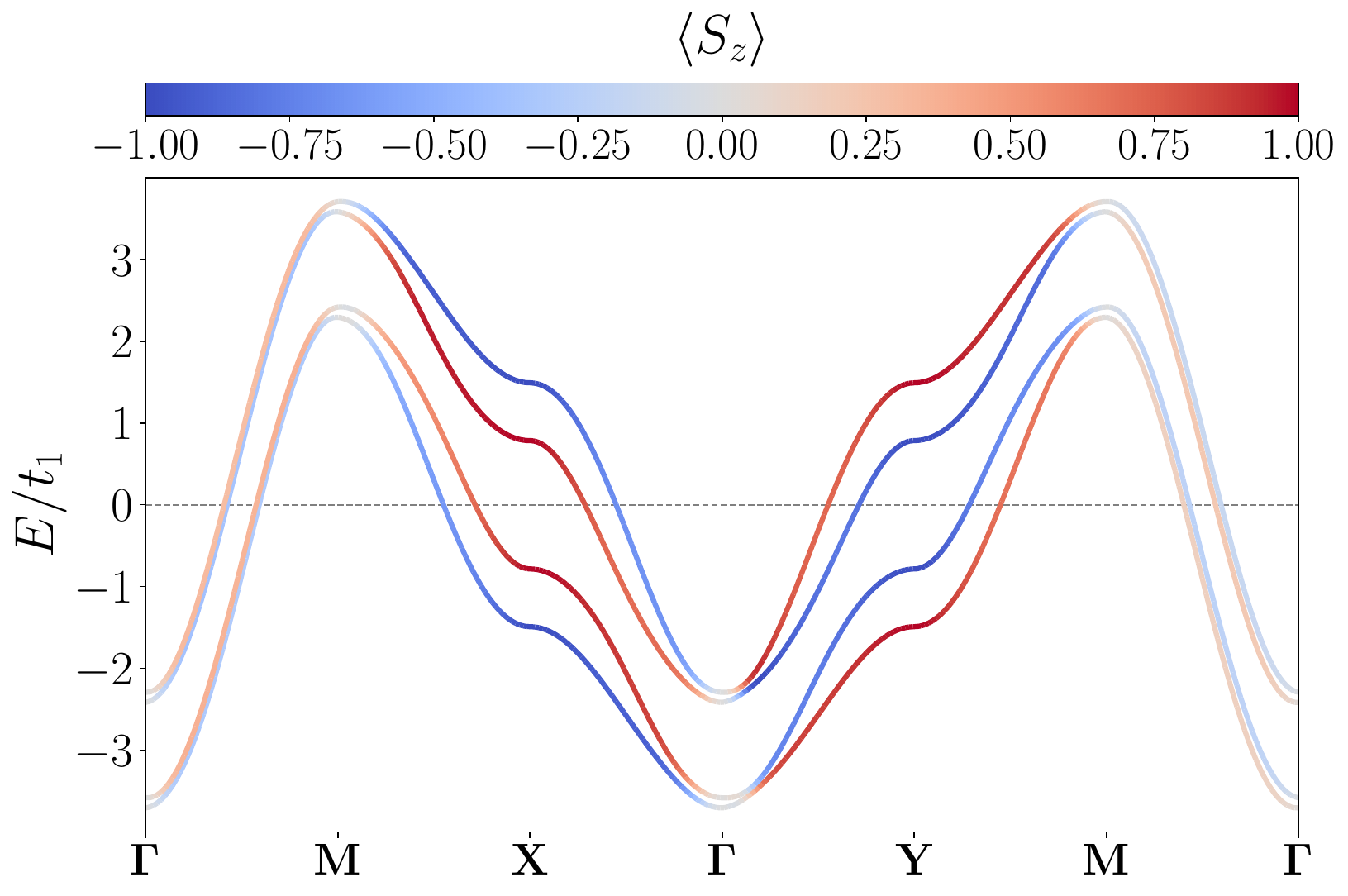}}
		\caption{The band dispersion for our system with intra-layer and inter-layer Rashba couplings and the following parameters: $t_2/t_1 =0.5$, $t_3/t_1 =0.5$, $J_{sd}S/t_1 =0.4$, $\lambda_1/t_1 =0.2$, $\lambda_2/t_1=0.1$, $\lambda_3/t_1=0.1$, $\phi=0$, and $\mu/t_1=0.0$. The color denotes the expectation value of spin in $z$ direction.}
		\label{fig:BS2}
	\end{figure} 	

	To achieve a nonzero AHE in this system, one option is to orient the magnetic moments within the plane. Probing the AHE would then require transport along the $z$-direction, which is impractical in our quasi-two-dimensional setup. An alternative approach to induce a nonzero AHE in our model is to apply an in-plane electric field. Both modifications of our model—either orienting the magnetic moments in-plane or applying an in-plane electric field—result in a finite magnetization~\cite{Optical}. Hence, we introduce the inter-layer Rashba coupling into our model that significantly impacts the electronic dispersion. To account for this inter-layer Rashba coupling, we introduce a tilted electric field, $\mathbf{E}_{\|} = (E \cos(\phi), E \sin(\phi), 0)$, where $\phi$ is the in-plane angle with respect to the x-direction. The inter-layer Rashba SOC Hamiltonian can be expressed as: ${H_R^{\text{inter}} = i \lambda^{\text{inter}} \sum_{i, \sigma} c_{1,i,\sigma}^{\dagger} (\boldsymbol{\sigma} \times \hat{\mathbf{z}})\cdot {\mathbf{E}}_{\|}} c_{2,i,\sigma}$. This model allows us to explore the intricate interplay between SOC and altermagnetic behavior in our system. The tight-binding model for the inter-layer Rashba Hamiltonian in our system is described by the following expression:
	\begin{equation}
		H^{\text{inter}}_R =  i \lambda_3 \sum_{i, \sigma} c^{\dagger}_{1,i, \sigma} (\sigma_x \sin(\phi) - \sigma_y \cos(\phi)) c_{2,i,\sigma} + h.c.,
	\end{equation}    
	where the magnitude of the in-plane electric field has been absorbed into the inter-layer Rashba coupling strength, $\lambda_3$. For simplicity, we consider $\phi=0$, meaning the in-plane electric field is oriented solely in the $x$-direction. The SOC between layers induces a gap in the electronic band structure and leads to the mixing of spin components across different layers, see Fig.~\ref{fig:BS2}.
	
\begin{figure}[t!]
		\centerline{\includegraphics[width=0.98\linewidth]{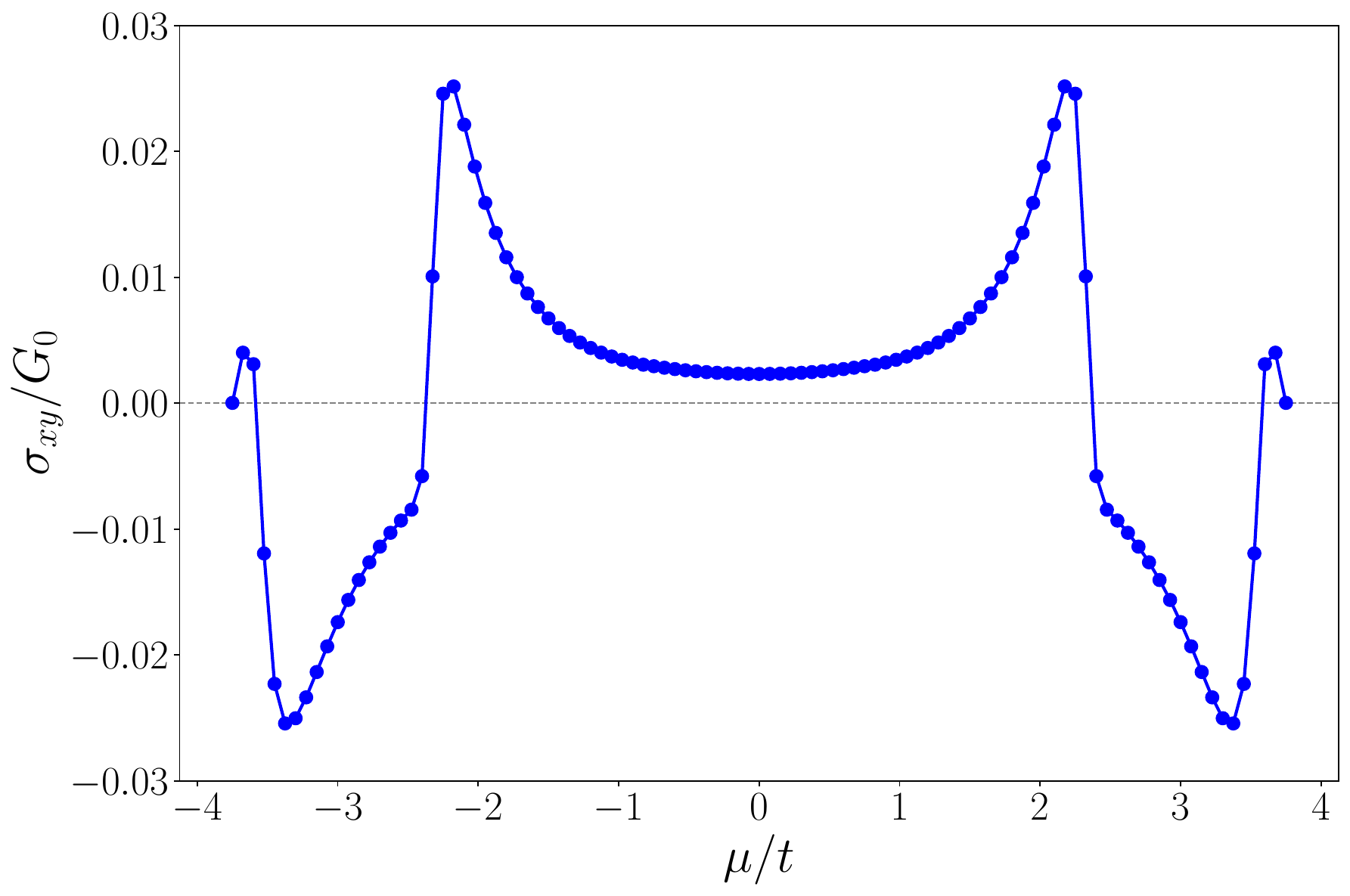}}
		\caption{The ratio of anomalous Hall conductivity to the conductance quantum $G_0$, as a function of the dimensionless chemical potential ($\mu/t_1$) for our system with the following parameters: $t_2/t_1 =0.5$, $t_3/t_1 =0.5$, ${J_{sd}S/t_1 =0.4}$, $\lambda_1/t_1 =0.2$, $\lambda_2/t_1=0.1$, $\lambda_3/t_1=0.1$, $\phi=0$, and $k_B T/t_1 = 0.01$.}
		\label{fig:AHC}
	\end{figure}

	Hence, we introduce the inter-layer Rashba coupling, which mixes the spin components of different layers. We find nonzero AHC when there is a nonzero Rashba SOC component along the alternmagnetic parameter order, $J_{sd}$, direction. This is consistent with the other minimal models presented for conventional, i.e., non-synthetic, altermagnets~\cite{models}. Figure~\ref{fig:AHC} illustrates the numerical calculation of AHC for our proposed synthetic altermagnet system with considering inter-layer and intra-layer Rashba coupling terms as the chemical potential varies. The magnitude of AHC of our synthetic altermagnet is in the range of 0.1-5 $\%$ of the conductance quantum $G_0=2 e^2/h$.

	\section{Conclusions}
	\label{sec_conclusions}
	
	In conclusion, our study introduces a novel approach to synthesizing altermagnets with properties analogous to conventional altermagnets, a new class of materials with significant implications for magneto-electronic research. Through comprehensive theoretical analysis and numerical calculations, we elucidate the electronic band structures, spin-polarized current, Berry curvatures, and AHC of our proposed synthetic altermagnets. Our findings show that the unique symmetry of altermagnetic electronic bands, along with their non-zero AHC, distinguishes them from conventional antiferromagnets. Recent work on twisted magnetic van der Waals bilayers, with their twist-induced Moiré patterns, further supports the relevance of our findings~\cite{vdW}, providing an alternative platform for exploring altermagnetism closely related to our synthetic altermagnet models. Additionally, recent studies on stacking configurations in bilayer altermagnetic systems based on symmetries broaden the range of candidate materials applicable in our approach to realizing two-dimensional altermagnets~\cite{Stacking, Stacking2}. These insights deepen our understanding of altermagnetism and highlight their potential for applications in spintronics and magneto-electronics. The unconventional spin-polarization and electronic properties of altermagnets present promising opportunities for advancing technologies based on their unique magnetic behaviors. We encourage further experimental investigation to validate and expand upon our theoretical predictions, paving the way for the practical realization and application of altermagnetic materials in advanced electronic devices.

	\section*{Acknowledgements}
	
	R. A. Duine is member of the D-ITP consortium, a program of the Netherlands Organisation for Scientific Research (NWO) that is funded by the Dutch Ministry of Education, Culture and Science (OCW). This work is part of the Fluid Spintronics research programme with project number 182.069, which
	is financed by the Dutch Research Council (NWO).
	
	%%%%%%%%%%%%%%%%%%%%%%%%%%%%%%%%%%%%%%%%%%%%%%%%%%%%%%%%%%%%%%%%%%%%%%%%%%%%%%%%%%%%%%%%%%%%%
	
	\appendix
	
	\section{Magnon Properties}
	\label{app_magnon}
	
\begin{figure}[t!]
		\centerline{\includegraphics[width=0.98\linewidth]{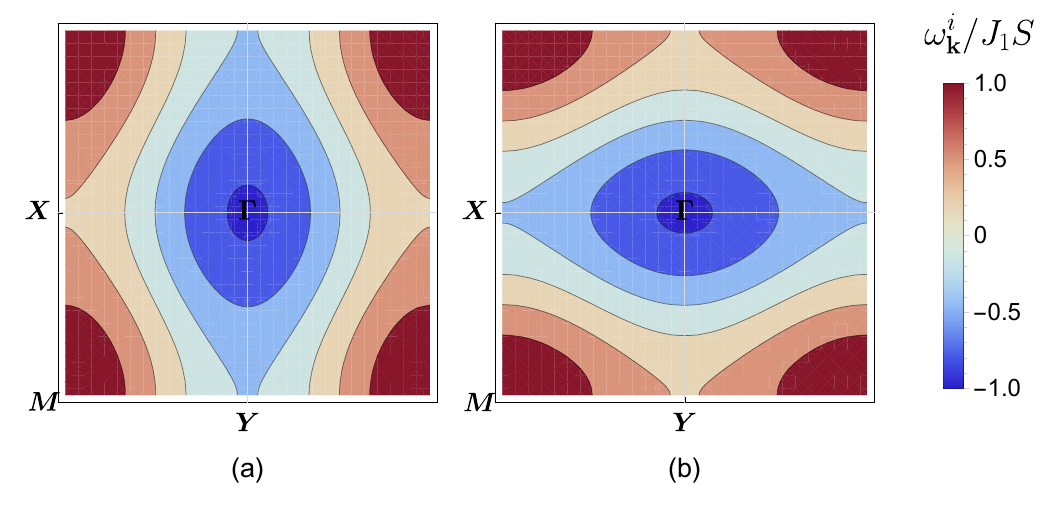}}
		\caption{The magnon modes, $\omega_{\mathbf{k}}^{i}/J_1 S$, with $J_2/J_1 =0.5$, and $J_3/J_1 =1.0$: a) $i=\alpha$, and b) $i=\beta$.}
		\label{fig:magmodes}
	\end{figure}

\begin{figure}[t!]
		\centerline{\includegraphics[width=0.98\linewidth]{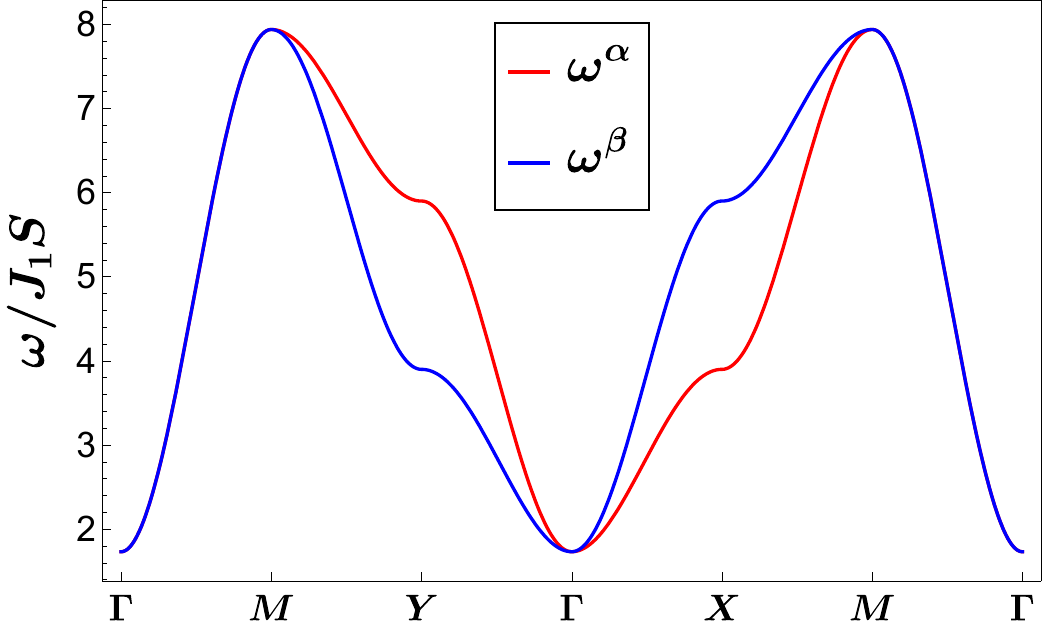}}
		\caption{The magnon spectra $\omega/J_1 S$. We used the parameters: $J_2/J_1 =0.5$, and $J_3/J_1 =1.0$. The color denotes different mode frequencies.}
		\label{Fig:magspectra}
	\end{figure}

	Here, we investigate the effects of anisotropic ordering of local orbitals on the spin-wave (magnon) spectra in reciprocal momentum space in agreement with the symmetry of our system. Specifically, we consider the leading interactions between localized spins, primarily focusing on spin-spin exchange interactions,
	\begin{eqnarray}
		H_m =&-&\sum_{\langle i,j \rangle_x} \left(J_{1} \left( \mathbf{S}^{A}_{i} \cdot \mathbf{S}^{A}_{j}\right) + J_{2} \left( \mathbf{S}^{B}_{i} \cdot \mathbf{S}^{B}_{j}\right)\right) \nonumber \\
		&-&\sum_{\langle i,j \rangle_y} \left(J_{2} \left( \mathbf{S}^{A}_{i} \cdot \mathbf{S}^{A}_{j}\right) + J_{1} \left( \mathbf{S}^{B}_{i} \cdot \mathbf{S}^{B}_{j}\right)\right) \nonumber \\
		&+& \sum_{\langle i,j \rangle}J_{3} \left( \mathbf{S}^{A}_{i} \cdot \mathbf{S}^{B}_{j}\right),
	\end{eqnarray}
	where $J_1$ and $J_2$ are the intra-layer ferromagnetic-like exchange coupling parameters, and $J_3$ denotes the strength of the inter-layer exchange interactions, coupling sites of opposite spin in an antiferromagnetic-like manner. We employ the Holstein-Primakoff transformation and subsequent Fourier transform to express the Hamiltonian up to bilinear terms in magnon operators, and then proceed to diagonalize it via using Bogoliubov transformation. The resulting magnon Hamiltonian is as follows,
	\begin{equation}
		H_m = \sum_{\mathbf{k}} \left[\omega_{\mathbf{k}}^{\alpha} \alpha^{\dagger}_{\mathbf{k}} \alpha_{\mathbf{k}} + \omega_{\mathbf{k}}^{\beta} \beta^{\dagger}_{\mathbf{k}} \beta_{\mathbf{k}} \right],
	\end{equation}
	where $\alpha_{\mathbf{k}}^{(\dagger)}$ and $\beta_{\mathbf{k}}^{(\dagger)}$ are the (creation) annihilation operators of magnon with momentum $\mathbf{k}$. The magnon frequencies are given by ${\omega_{\mathbf{k}}^{\alpha (\beta)} = \pm \gamma_1 + \sqrt{\gamma_2^2 - \gamma_3^2}}$ with
	\begin{subequations}
		\begin{flalign}
			&\gamma_1 = (J_2 - J_1) S (\cos(k_x a) - \cos(k_y a)),     \\
			&\gamma_2 = -(J_1+J_2) S  (\cos(k_x a)+\cos(k_y a)) + 2 S \sum_{i=1}^3 J_i,  \\
			&\gamma_3= J_3 S.
		\end{flalign}
	\end{subequations}
	These parameters capture the intra-layer ferromagnetic-like exchange interactions ($J_1$ and $J_2$) and the inter-layer antiferromagnetic-like exchange interaction ($J_3$).   
	The altermagnet magnon bands can exhibit chirality and carry spin current similar to ferromagnets, while possessing linear spectra around the $\Gamma$ point in momentum space akin to antiferromagnets~\cite{Sinova}. In our system, the anisotropic coupling strength in different directions leads to unconventional $d$-wave symmetry magnon bands in momentum space, see Figs.~\ref{fig:magmodes} and~\ref{Fig:magspectra}. 

	Figure~\ref{Fig:magspectra} shows split magnon bands with opposite chirality and non-zero spin expectation values.
	However, the magnon dispersion of our model is not linear around the $\Gamma$ point due to the lack of translational symmetry in the $z$-direction. When stacking layers with alternating magnetic orders, the magnon dispersion becomes linear around the degenerate point, similar to antiferromagnets.The magnon band splitting in our system demonstrates an unconventional symmetry resembling electronic spin splitting, as described in Eq.~\ref{eq:gap}, and can be expressed in momentum space as follows:
	\begin{equation}
		\Delta_{m} = 2 S (J_2 - J_1) (\cos(k_x a) - \cos(k_y a)).
	\end{equation} 
	%
	
	%%%%%%%%%%%%%%%%%%%%%%%%%%%%%%%%%%%%%%%%%%%%%%%%%%%%%%%%%%%%%%%%%%%%%%%%%%%%%%%%%%%%%%%%%%%%%

\end{document}